\documentclass{article}
\usepackage[utf8]{inputenc}

\usepackage{authblk}

\usepackage{fullpage}
\usepackage{palatino}
\usepackage{accents}
\usepackage{xcolor}
\usepackage{longtable}

\usepackage{tikz}
\usetikzlibrary{shapes.multipart}
\usetikzlibrary{decorations.pathreplacing}

\usepackage[breaklinks=true]{hyperref}
\usepackage{cleveref}
\definecolor{webgreen}{rgb}{0,.5,0}
\definecolor{webblue}{rgb}{0,0,.5}
\hypersetup{
    colorlinks,
    linkcolor={green!50!black},
    citecolor={green!50!black},
    urlcolor={green!50!black},
    filecolor={green!50!black},
}

\usepackage{enumitem,amssymb}
\newlist{todolist}{itemize}{2}
\setlist[todolist]{label=$\square$}
\usepackage{mdframed}

\title{A Guide for New Program Committee Members at Theoretical Computer Science Conferences\footnote{Feedback and suggestions are welcome and should be sent to \href{mailto:c.schaffner@uva.nl?subject=Feedback about PC guide}{\texttt{c.schaffner@uva.nl}}.}}
\author[1]{Yfke Dulek}
\author[1]{Stacey Jeffery}
\author[1]{Christian Majenz}
\author[2]{\linebreak Christian Schaffner}
\author[2]{Florian Speelman}
\author[1,2]{Ronald de Wolf}
\affil[1]{CWI, QuSoft}
\affil[2]{University of Amsterdam, QuSoft}
\date{\today}

\begin{document}

\maketitle

\begin{abstract}
In theoretical computer science, conferences play an important role in the scientific process. The decisions whether to accept or reject articles is taken by the program committee (PC) members. Serving on a PC for the first time can be a daunting experience. This guide will help new program-committee members to understand how the system works, and provide useful tips and guidelines. It discusses every phase of the paper-selection process, and the tasks associated to it.
\end{abstract}

\section*{Introduction and Overview}
\subsection*{Scientific method and the role of conferences}
One of the core principles of modern scientific research is \emph{peer review}. New research results are written up as articles and submitted to journals or conferences where they are reviewed by (usually multiple anonymous) experts in the field. These experts judge whether the article fulfills the scientific standards and fits the publication venue. Only after an article has passed this quality check by peers, it is considered \emph{published} and becomes part of the scientific literature.

Publication cultures differ vastly between different research fields. In computer science, it is common that research articles are first submitted to conferences, and only a subset of those articles are worked out to "full versions" and later submitted to journals. In sharp contrast, the standard practice in mathematics and physics is to submit new research articles directly to journals. 

These differences in publication cultures lead to a very different understanding and experience of the role conferences play in this process. In computer science, conferences and their published proceedings are the main venues for publication and therefore very important. Researchers have quite a clear picture in their head about the "ranking" or "tier" of all conferences they submit to. The more prestigious conferences in theoretical computer science (TCS) like FOCS and STOC tend to receive hundreds of submissions, and have acceptance rates of around 30\%, and the best conferences specialized in a particular subfield tend to have even lower acceptance rates. It is important for the career of junior researchers (PhD students and postdocs) in TCS to have articles published at good conferences, and the task of giving a conference presentation is often offered to these junior researchers in order to give them the opportunity to make their first contacts with the research community.

For a long time, it has been common practice in some subfields of physics to make new research articles immediately available to the whole research community by putting them as \href{https://en.wikipedia.org/wiki/List_of_academic_preprint_repositories}{pre-prints on servers} like \href{https://arxiv.org}{arxiv.org}. It is great to see that almost all of computer science has adopted this practice by now. The two main effects are the following: On the one hand, it allows the research community to advance faster (without having to wait for the outcomes of the peer-review process). On the other hand, because all new research articles are freely available from these pre-print servers, it allows the community to \href{https://en.wikipedia.org/wiki/Academic_journal_publishing_reform}{move away from commercial publishers}, making publicly funded research available to the whole world, and supporting new open-access publication models such as \href{https://en.wikipedia.org/wiki/Overlay_journal}{overlay journals}.


\subsection*{What is a program committee (PC)?}
The reviewing process at a conference is coordinated by a program committee (PC). The members of this committee are themselves scientists that are active in the field of the conference. They are responsible for deciding which submissions will be accepted for a presentation and/or publication at the conference. Depending on the topic of your submission, a PC member can act as a reviewer themselves, or they can find a so-called subreviewer to give them an expert opinion about your submission.\footnote{During the later stages of your PhD or beginning stages of a postdoc, you may have been contacted by someone requesting such a subreview -- that person was serving on a PC.}

Although some PC members stay on for multiple years, a new committee is generally formed for every new edition of a conference. The responsibilities of a PC member therefore center around a period of a few months in a single year.

Note that TCS has a somewhat different conference process than other areas of computer science. A PC member in TCS roughly corresponds to "area chair" in conferences in other areas such as machine learning. 

\subsection*{Timeline}
Most PCs follow a similar timeline, starting at the moment you receive your invitation, and ending at the conference itself. Most of the work is concentrated around the bidding, review, and discussion phase.

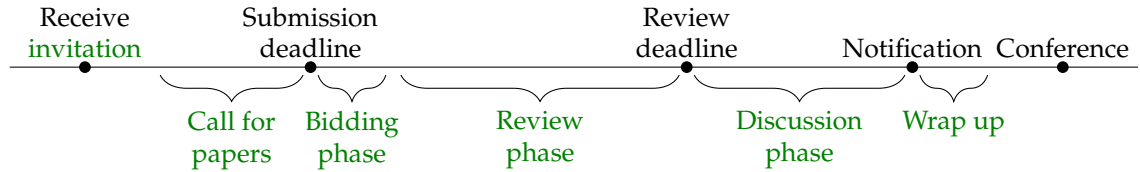
\begin{center}
\begin{tikzpicture}[every text node part/.style={align=center}]
\draw (0,0) -- (15,0);
\draw[fill=black] (1,0) circle (2pt);
\node[anchor=south] at (1,0) {Receive\\ \hyperref[sec:invitation]{invitation}};

\draw[decorate,decoration={brace,amplitude=10pt, mirror},yshift=-3pt]
(2,0) -- (3.9,0) node [midway,anchor=north,yshift=-10pt] {\hyperref[sec:CFP]{Call for}\\ \hyperref[sec:CFP]{papers}};

\draw[fill=black] (4,0) circle (2pt);
\node[anchor=south] at (4,0) {Submission \\ deadline};

\draw[decorate,decoration={brace,amplitude=10pt, mirror},yshift=-3pt]
(4.1,0) -- (5,0) node [midway,anchor=north,yshift=-10pt] {\hyperref[sec:bidding]{Bidding}\\ \hyperref[sec:bidding]{phase}};

\draw[decorate,decoration={brace,amplitude=10pt, mirror},yshift=-3pt]
(5.2,0) -- (8.9,0) node [midway,anchor=north,yshift=-10pt] {\hyperref[sec:review]{Review}\\ \hyperref[sec:review]{phase}};

\draw[fill=black] (9,0) circle (2pt);
\node[anchor=south] at (9,0) {Review \\ deadline};

\draw[decorate,decoration={brace,amplitude=10pt, mirror},yshift=-3pt]
(9.1,0) -- (11.9,0) node [midway,anchor=north,yshift=-10pt] {\hyperref[sec:discussion]{Discussion}\\ \hyperref[sec:discussion]{phase}};

\draw[fill=black] (12,0) circle (2pt);
\node[anchor=south] at (12,0) {Notification};

\draw[decorate,decoration={brace,amplitude=10pt, mirror},yshift=-3pt]
(12.1,0) -- (13,0) node [midway,anchor=north,yshift=-10pt] {\hyperref[sec:wrapup]{Wrap up}};

\draw[fill=black] (14,0) circle (2pt);
\node[anchor=south] at (14,0) {Conference};
\end{tikzpicture}
\end{center}
Click on the links to learn more about each phase.

\subsection*{Abbreviations and terminology}
\begin{longtable}{p{4cm} l p{11cm}}
    TCS & \qquad & Theoretical computer science\\
    PC & & Program committee (selects accepted papers)\\
    PC chair & & Program-committee chair (selects and leads PC), role is often fulfilled by two people\\
    SC & & Steering committee (selects PC chair, invited speakers, and is involved in organizing the conference)\\
    Local organizing committee & & Takes care of the more practical details of organizing the conference (venue, food, website, etc.)\\
    Subreviewer & & Provides expert review on a submission (asked by a PC member)\\
    COI & & \hyperref[sec:coi]{Conflict of interest}\\
    Review score(s) & & Discrete value(s) attached to a review, indicating (scientific/editorial) quality of submission, confidence of the reviewer, etc.\\
    EasyChair, HotCRP & & Commonly used online tools for managing the review process.
\end{longtable}

\section{Invitation}\label{sec:invitation}
Before the call for submissions is published, the program committee chair has to assemble the members of the program committee. At this point in time, you may receive an invitation asking you to serve on the program committee for that year's edition of the conference.

\begin{mdframed}
During this phase:
\begin{todolist}
\item You gather information to make a decision about whether you want to join the PC.
\item You respond to the PC chair to accept or decline.
\end{todolist}
\end{mdframed}

\subsection{How to get invited to a PC?} 
Usually it is up to the PC chair(s) to select the PC members, so you have to be on this person's radar. PC chairs usually try to select a committee that covers all of the relevant research areas, is diverse in terms of people (in TCS, there is usually a higher demand for women and other minorities), and a good mix of more senior but also eager junior members. Senior members tend to provide more knowledgeable and more ``expert'' reviews, taking into account the broad picture, and all previous research. Junior members usually have more time and spend more effort to provide more extensive reviews, including closer reading and proof-checking.

The following actions increase your chances to get invited to a PC:
\begin{itemize}
\item Do good research, publish in good venues, and give good talks at conferences.
\item Provide good subreviews for conferences and journals, and be reliable as a subreviewer, e.g., by replying timely to review requests, stick to deadlines etc.
\item Happen to have a personal and/or professional connection with the PC chair. The PC chair often looks back on PCs which they have been on themselves, and checks the other PC members' performances. Reliability is a very important factor in their decision.
\item Get recommended by somebody (usually more senior) who declines to be on the PC. You could actively let your close colleagues/collaborators/supervisors know that you are looking for this opportunity.
\item At some conferences, the steering committee keeps a list of ``promising'' young people who should be invited to serve on PCs soon. It is a bit pushy, but it might actually work to approach these members and offer to help on a next PC.
\end{itemize}

\subsection{Accept or reject the invitation}
Think carefully before accepting the invitation to join a program committee. Make sure to have gained some experience as a subreviewer before joining your first program committee, so you have some idea of the process and the expected time investment.

\paragraph{Advantages} of serving on a program committee:
\begin{itemize}
\item You actively contribute to a core part of the scientific process (namely to peer-review): serving on a PC is a service to the community and to science in general.
\item PC memberships increase your scientific credibility and visibility. It will be noticed by influential people if you are doing a good job on a PC.
\item You have (partial, but possibly quite a lot of) power to decide what is accepted to a conference. Evidently, with this power comes the responsibility to use it for the good of science and not for selfish motives.
\item You get an inside view of how the conference-review-and-selection process actually works (e.g. how much randomness is involved, what PC members care about in the discussion etc.), and can use this knowledge when preparing submissions for future conferences.
\item You get a sneak preview of the latest and greatest research results in the field.
\item You are externally motivated to read some interesting papers in depth.
\end{itemize}

\paragraph{Disadvantages} of serving on a program committee:
\begin{itemize}
\item It costs a lot of time, energy and (potentially) nerves. Check carefully at what dates most of the work will be required (your highest workload will be right after the submission deadline, then it will be up to you how to balance the workload during the review phase, and then potentially high during the discussion phase), how many submissions you are expected to review, how close it will be to your core expertise etc. Expect to spend at least 2 to 3 hours \emph{on average} on each paper you assign to yourself, excluding the discussion phase. As a junior PC member, you are likely to spend more time on the task.
\item If you do a bad job (like not providing reviews on time, not participating in the discussion, having vastly differing opinions), the community will notice, starting with the PC chair, and the other PC members you interacted with, and this will decrease your chances for being asked to join another PC, and, more generally, will negatively affect your colleagues' opinion of you.
\item You are forced to read some bad submissions.
\end{itemize}

\paragraph{Warning:} Only consider serious offers, for conferences that you know and where you would consider submitting your own work. You should never have to pay to be part of a PC. Be wary of fake or spam invitations; when in doubt, ask a more senior member of your group for their opinion.

\section{Call for papers}\label{sec:CFP}
Once you accept the invitation, the PC chair draw up the ``call for papers'', which will include your name and affiliation as a PC member. Sometimes a picture is included as well. The call for paper specifies the page limit, formatting, and whether simultaneous submission or submission of previously published papers are allowed. As PC member, you might voice your opinion on these issues to the PC chair. The call for papers also contains the deadlines and deadline for notification, so you might want to check if those dates make sense. Sometimes there will be an abstract submission deadline followed by a full paper submission deadline.

During the rest of this phase, authors will prepare their manuscripts for submissions, so you do not have to do much. For most conferences, you are allowed to submit your own work as well (as an author). There may be limitations: for example, there may be a maximum number of allowed submissions, or your submission(s) may be held to higher standards.

\begin{mdframed}
During this phase:
\begin{todolist}
\item You double-check that your name is spelled correctly and the affiliation says what you want.
\item You read the call for papers carefully and make sure your area of expertise is correctly represented.
\end{todolist}
\end{mdframed}

\section{Bidding phase}\label{sec:bidding}
Within a few days after the submission deadline, the PC chair will activate the bidding phase, where all PC members see a (usually very long and possibly intimidating) list of all submissions. You have a few days to indicate your preference for reviewing particular submissions. The goal is for the PC chair to assign three PC members to each paper (or more, typically five, for papers submitted by PC members), so you can calculate approximately how many papers will end up in your responsibility.

\begin{mdframed}
During this phase:
\begin{todolist}
\item You look at all submissions (titles, abstracts, and an occasional closer look).
\item You ``bid'' on submissions to express your interest in being responsible for reviewing them.
\item You indicate any \hyperref[sec:coi]{conflicts of interest (COIs)}.
\end{todolist}
\end{mdframed}

Bidding is often done by assigning points to each submission, for example ranging from -20 (``I really don't want to review this'') to +20 (``please please assign me to this one''). The system works best if you use the whole spectrum of points, and not just the two extremes.

Go through the list of submissions (looking at titles and abstracts, sometimes downloading the PDF and having a closer look), and give clear indication of which submission you would actually like to review, and maybe even more importantly, which you do not want to review. If there is a search functionality, you can search for keywords (e.g., ``quantum'' or ``cryptanalysis'') in order to make a rough distinction between your favourite topics and those you have no idea about.

The most useful thing you can contribute are \emph{high-confidence reviews}, either by yourself or by people you can ask as \hyperref[sec:subreviewers]{subreviewers}. If you find there are not many papers in your area(s) of expertise, you can also bid on papers on topics that you have close colleagues working on, so you will know some subreviewers, and they are likely to accept your request. However, keep in mind that in the end, you are responsible for making sure a proper review is submitted, so you have to be able to form an opinion if necessary. You should also be able to contribute meaningfully to the discussion phase.

You can (and should) bid on any paper that you think you can confidently review. This is not necessarily limited to papers under your expertise. For example, if there is a submission on Basket Weaving, and you are confident that Basket Weaving does not fall under the topics of the conference, you can confidently write a rejecting review without having any expertise in Basket Weaving.

\subsection{Conflicts of Interest}\label{sec:coi}
During the bidding phase, you are also asked to indicate possible conflicts of interest (COI) that have not already been caught by the system. COIs are often handled slightly differently from conference to conference, the PC chair will give you clear instructions about this. If in doubt, double-check with the chair. At pretty much any conference, submissions (co-)authored by
\begin{itemize}
\item yourself,
\item a person from the same institute as yours,
\item your (former) PhD student(s) and recent postdocs, or (former) PhD or postdoc advisor,
\item family members or close personal relationships
\end{itemize}
should be indicated (by you) as COIs, as you cannot be impartial while judging these submissions. In general, mark any submission where you feel unable to be impartial as COI, e.g.\ because of ``previous history'' you have had with an author or a particular submission. If in doubt, discuss with the chair.

Once you mark a submission as COI, it will completely disappear from your view. This can be weird, because other PC members still see the submission (and its discussion), and you might hear them discussing a certain number of submissions in a particular state, and those numbers might not add up for you because of COIs. Basically, only the PC chair has the complete view of the whole process. PC chairs are often not allowed to submit anything themselves, as this complicates things even further (a PC co-chair would have to handle the PC chair's submissions).

Some conferences distinguish between ``hard COI'' and ``soft COI''. We described ``hard COI'' above, meaning that you should have completely no involvement in discussion of the paper, and it should disappear from your view. A ``soft COI'' means that you have some interaction with the paper (for example, you have a competing paper, or maybe you had some old collaboration or connection with the author that does not rise to a hard COI). In this case you should declare your COI in the comments to the PC, so people are aware of it.

\subsection{Ethical dilemmas}\label{sec:ethical}
At conferences with anonymous submissions, refrain from googling the title of the paper in order to find out author names. You can be more impartial if you do not know (exactly) who wrote the submission.

You are likely to encounter submissions that are very related to your own work (such as your own submissions to this or other conferences, almost finished work in progress, ideas you had previously, work of your students, etc.). On the one hand, your closeness to the subject area makes you the ideal referee for such submissions. On the other hand, your closeness to the subject makes it impossible or very hard for you to be impartial. 
It is important to keep in mind that as a PC member, you are bound to \textbf{confidentiality, and you are not allowed to use any information or research results you encounter in the review process for your own research until the submission appears publicly}. In practice, you can hardly make your brain split in halves and only think about your own work in one half, and do the PC work in the other half. In most cases, the submitted results will appear publicly (on a pre-print server) not long after submission and most definitely after acceptance. You have several options:
\begin{itemize}
\item The easiest way out is to declare a COI, the submission will disappear from your worldview, and you try to forget you have ever seen it (at least until it pops up on some pre-print server).
\item If you feel that you are able to give an (almost) impartial review (in particular if there are probably no other qualified referees available), you should do so, but clearly indicate in the ``comments to the PC'' what your potential COI is. The situation is easier if your review is positive, because a negative review can be seen as you shooting down submissions competing with your own submission or interests. In case of a negative review, try to give good technical reasons, and possibly even comparisons to your own work.
\end{itemize}

Unfortunately, it has happened before that paper X is submitted, a PC member gets it for review, and then posts a paper X' with very similar result, and to add insult to injury, the paper X gets rejected and now it has been scooped by X'. Such episodes should increase your motivation to post your results publicly right after the time of submission.

\section{Review phase}\label{sec:review}
After all the PC members have indicated their reviewing preferences, usually 5-7 days after the submission deadline, the PC chair runs some optimization algorithm that assigns papers to reviewers. There might also be a fair amount of manual adjustments. Then the reviewing phase starts, and generally lasts around a month. At the end of the reviewing phase, all your assigned submissions should have review score(s) and a review text. There is also an option for comments to the PC (visible to other PC members, but not the authors of the submission), or comments to the PC chair only. In addition to these three fields, there may be specific questions (e.g., about editorial quality of the submission) which can be free form or a numerical scale.

\begin{mdframed}
During this phase:
\begin{todolist}
\item You familiarize yourself further with the submissions that were assigned to you.
\item You review several submissions yourself.
\item You contact external subreviewers to review the other submissions under your responsibility.
\item You notify the PC chair of any bugs you or your subreviewers find in otherwise good submissions; they might give the authors a chance to rectify.
\item You remind your subreviewers that their reviews are due.
\item You edit and submit all reviews/scores before the deadline.
\item You allocate time for last-minute reviews in case one of your subreviewers fails to deliver.
\end{todolist}
\end{mdframed}

As soon as you receive your pile of papers, download them and familiarize yourself a little better with them. This pile of paper will entertain you for a substantial part of the following month. For each submission, get a feeling of the following aspects:
\begin{itemize}
\item How likely it is that you can provide a high-confidence review yourself?
\item How long would it take you to review this submission? You probably already have an idea of how long it takes you to conference-review a paper. As a guideline, a minimal review of a submission with chances of acceptance will take you at least half a work day.
\item Can you think of a person (or two) who could ideally subreview this submission, in case you cannot review it yourself?
\end{itemize}
Decide how many submissions of your pile you have time to review yourself, and how many you want to outsource to subreviewers. This decision depends on the amount of time and energy you have available, but generally PC members review about half of their assigned papers themselves. Some types of papers are suitable for reviewing yourself: papers you have already read, papers you anyway want to read, papers close to your expertise, papers that are easy to dismiss (because they seem out of scope, very messily written, claim that P=NP, etc.).

You can also ask for informal opinions from colleagues who may not have time to give a full subreview, but are well positioned to comment on the context of certain results. In that case, indicate that you have done so in the comments to the PC. Collecting such an informal opinion can also be an option if you are having trouble finding a subreviewer and the deadline is getting close, as someone might be willing to respond with an informal opinion, especially on some specific aspect of the paper, who would not otherwise commit to writing a full review in the short timeframe. 

\subsection{Writing reviews yourself}
A must-read is Ian Parburry's (very old, but timeless) \href{https://jmlr.csail.mit.edu/reviewing-papers/p92-parberry.pdf}{Guide for New Referees in Theoretical Computer Science}. Remember that the goal is to provide \emph{high-confidence reviews}! A good conference review is not too long, and it summarizes the main contributions of the submissions which also benefits the other PC members, starting off the discussion. In case there is a rebuttal phase, clearly formulate questions that you would like the authors to answer.

For conferences without proceedings, do not spend too much time giving feedback on the presentation, because the goal is not to produce a perfectly polished piece of writing in the proceedings. If a paper is badly written, especially if there are (many) typos in the mathematics, this can of course still affect the final score.

As PC member, you should try to focus on what the central contributions of the paper are, rather than finding flaws in it. If a paper has a novel idea and would benefit the community to be aware of, then it might be worth accepting even if it has typos and some sloppy references. On the other hand, if the main result is boring, it does not matter how solid the execution is.

Give rather extreme scores if you love or dislike submissions. Grades tend to be distributed normally around the mean, so a greater variance is appreciated. You can always adjust your grades later. Do not be afraid of indicating high confidence for submissions in your area, you \emph{actually are} one of the experts in the field. Similarly, indicate (very) low confidence for submissions far outside of your expertise where your review is just a wild guess, try to avoid those anyways.

The top submissions (clear accepts) and bottom submissions (clear rejects) are always easily identified, but the big bulk of the submissions will be in between. Be aware that the border between acceptance and rejection is not the middle score, but rather (usually) ``weak accept'': some submissions with an average ``weak accept'' score may get in, most will not. Often, a submission needs at least one very positive opinion in order to get accepted.

\subsection{Managing subreviewers}\label{sec:subreviewers}
\textbf{Do not wait for longer than a day with contacting potential subreviewers for a submission.} Often there is one ``obvious'' reviewer for a paper, and other PC members will attempt to ask this person as well. 

Send along the review guidelines (and COI policy) to subreviewers. Often you can simply point them to a URL for this information. Give them a deadline which is 2-3 days before your own deadline, in case people are unreliable or something else goes wrong. This way, you have time to read through their review before entering it into the system, possibly adjusting the grades to what is expected, or mix in some of your own opinions. Some subreviewers fail to send in a review altogether, so reserve some time close to the deadline to do some last-minute reviews yourself.

Depending on the type of conference and the size of your network, potential subreviewers might decline or never even respond. You can decrease this probability by:
\begin{itemize}
\item sending out your requests as soon as possible, so that the subreviewer did not already agree to review for another PC member;
\item following up on your request if you don't get a response to your request within approximately three days. Easychair (and other automated) emails sometimes end up in people's spam folder or are otherwise ignored. You can use your own email to follow up, asking them to respond to the referee request;
\item asking potential subreviewers ``informally'' (via email or in-person) before sending out the official referee request. This partially resolves the spam-folder issue, but requires you to keep track of more communication channels;
\item adding a personalised message to the review request, rather than the standard message provided by the system;
\item asking people that you know personally, or for whom you have reviewed in the past.
\end{itemize}

If you do not know any potential subreviewers with expertise in the topic of a given submission, you can look through the list of references for recent relevant papers and ask one of their authors. If the submission is a follow-up to a specific paper by different authors, you can ask one of those. You can also look at the acknowledgement section, but those people could be ``too close'' to the submission. Avoid asking subreviewers from the same institute as the authors (this is a COI), other PC members, and of course the authors themselves. If the submissions are anonymous, you cannot avoid potentially doing so. In case that happens, the subreviewer will decline because of a COI.

Ask your potential subreviewer for an explicit confirmation that they can commit to doing the review. If they do not respond, not even after a reminder, continue looking for another reviewer. Do not wait too long with this. You could even prepare a list of back-up subreviewers to ask, so that you can act quickly in case a request is declined or ignored.

To the people that did respond and agree to review, send a reminder a few days before the deadline you have given them, and again if the review has not come in after the deadline is passed. Here is a template for a reminder email:

\begin{verbatim}
Dear X,

On [date] I sent you (via EasyChair) a referee request for the [conference
name] submission [name of submission], which you kindly accepted.

I didn't get a reply yet. It is of course fine if you don't want to do it,
but could you please let me know either way, so that I know whether I have
to look for someone else?

Best regards,
[you]
\end{verbatim}

 Decide beforehand when to give up and switch to doing the review yourself. Usually, if one post-deadline reminder does not work, it is time to give up. Remember to reserve time to quickly review one or two papers yourself in case one of your subreviewers goes missing.

When a subreviewer submits their review, do not just copy-paste it into the reviewing system, but instead:
\begin{itemize}
\item Have a look at (at least) the introduction of the submission yourself. If the submission has only external reviewers and they are all positive, it may be wise to read the entire paper, so that at least one PC member has had a close look at it.
\item Edit and error-correct the review itself if needed. Doing so will also protect the privacy of your subreviewer: for example, it removes grammar mistakes that are typical to native speakers of specific languages.
\item Recalibrate the referee's score(s) to ensure consistency among the submissions under your responsibility. If the acceptance rate of the conference is 20\%, then roughly 20\% of the submissions in your pile should receive an accepting score. But beware: this is by no means a hard rule, because your batch is biased: your area of expertise may be especially ``hot'' and fast-paced this year, you may have identified many obvious rejects during bidding, etc. Often, recalibrating means toning over-enthusiastic scores down a notch to bring them in line with other scores in your batch.
\end{itemize}
In case you make any changes to the text or scores of the subreview, you should specify and explain these changes concisely in the "comments to the PC" field.

\subsection{Submissions by PC members (like yourself)}
Submissions by PC members often get more reviewers assigned (say, five instead of three), which usually means raising the difficulty to get accepted. The standards applied for those submissions are often a bit stricter than for other submissions. These higher standards can be a reason for preferring a competing submission of similar quality.

Often, there is a cap (usually around two) on the number of papers that a PC member is allowed to submit. This cap may interfere with your desired research output, especially if you are in the postdoc phase of your career. However, the benefits of serving on a PC of a relevant conference usually outweigh this drawback.

\section{Discussion phase}\label{sec:discussion}
You will not be able to see other PC member's reviews until you have submitted your own. In most reviewing systems, there is the option of putting submissions on a personal ``watchlist'', which means that you get notified if there is a new comment in the discussion. Make sure all submissions on your pile (and others you are interested in) are on your watchlist. This does not always happen automatically.

\begin{mdframed}
During this phase:
\begin{todolist}
\item You read the other reviews on the submissions in your pile, and potentially revise your score.
\item You start and contribute to discussions on the papers in your pile, with the goal of clearing up any questions and reaching a consensus about acceptance or rejection.
\item You may be asked (or be interested) to contribute to certain discussions outside your pile. Proactive contributions on top of your own pile are highly appreciated in this phase.
\item You might give feedback to, or ask additional input from, your subreviewers.
\item In case the conference allows rebuttals, you read and consider the authors' rebuttals.
\item You might vote or attend a physical PC meeting for the final decisions.
\item You polish (or revise) reviews at the end of this phase, before they are being sent to the authors.
\end{todolist}
\end{mdframed}

Once the discussion phase has started, read carefully through the reviews of the other PC members (or their subreviewers), and check the scores they have given. You might want to look up names of subreviewers that are unfamiliar to you. Immediately start off the discussion in case you disagree with somebody else's opinion. If another review provides you with new and relevant information that you had missed (for example, another reviewer finds an error in a proof that cannot easily be fixed), you may decide to revise your own score based on that new information. That will help the PC chair to identify submissions with a low (or high) score.

If you have limited time, focus on the borderline cases, as the good and bad submissions will ``automatically'' get accepted or rejected. Contribute to the discussion where it is necessary and useful.

In case of very similar submissions, there might be comparative discussion boards, where two or more submissions are directly compared to each other, and all PC members from the individual submissions can participate (except if they have COIs with any of the involved submissions). In these comparison discussions, expert opinions of people knowing the details of multiple submissions are highly appreciated. Again, such a joint discussion can be rather weird (like a game of \href{https://en.wikipedia.org/wiki/Mafia_%28party_game%29}{Werewolf}) if you or others have COI with one of these papers. If this is the case, you should exclude yourself from the discussion as much as possible. You may even consider avoiding both submissions if you can already foresee a possible merger during the bidding phase, to prevent an \hyperref[sec:ethical]{ethical dilemma}.

Sometimes, the PC chair will ask you explicitly to have a look at another submission (on top of those that you have already reviewed). Read at least the title, abstract, and introduction to form a general opinion before looking at the other reviews.

During the discussion phase, you do not have to limit yourself to the papers initially assigned to you. You should look for other papers where discussion is incomplete (i.e., reviewers have presented conflicting opinions and have not responded to each other), you have relevant expertise and/or the ranking looks borderline.  But also look for misclassified papers that are far from the borderline. You can also invite in other PC members to join the discussion. Try to watch out for papers in areas that are not "fashionable" or are not well represented in the PC, and make sure they do not fall through the cracks due to lack of attention. \textbf{In general you should be proactive and not assume that other people will take care of things that seem wrong.}

If you believe it has added value, you could ask the PC chair for the permission to forward the other reviews (without scores or names) to your subreviewer, and ask them for their input to the discussion. In particular, this could be useful for more junior subreviewers, also to give them a better idea how this whole reviewing process actually works, and what other (possibly much more senior) experts in the field have to say about the submission that they analyzed very carefully. 

\subsection{Rebuttal phase}
Some conferences have a rebuttal phase, where the authors can read and respond to (``rebut'')  reviews on their work. Depending on the settings, authors might not see the numerical scores, and no (final) decision has been made about acceptance or rejection at this point. The discussion phase continues after the rebuttal has been received.

Read through the authors' rebuttal. Do they address the concerns you had? If you believe it has added value, you might forward the authors' rebuttal to your subreviewer and ask for input to the discussion.

Instead of an explicit rebuttal phase, some conferences allow PC members to send questions to the authors via the PC chair, usually when there are narrow technical concerns about correctness, or other factual points.

\subsection{Decision phase}
As the discussion phase draws to a close, it is time to make decisions. Normally, the decisions about which submissions to accept will be initiated by the PC chair. Based on the given scores (possibly weighted with the confidence score), the PC chair will change the ``status'' of a submission to ``preliminary accept'', or ``preliminary reject''. If nobody speaks up in the discussion for a week or so, these decisions become final.

Clear accepts and clear rejects are usually easy to handle. Most time will be spent on discussing submissions with diverging scores from different reviewers, and ``borderline'' papers.

About a week before the date of notification, there are usually quite a bunch of (say 40) submissions left that are all competing for the remaining (say 20) spots. In this phase, pretty much any of the submissions in the running could make it in or not. If you want to, you can have a lot of influence in this phase by voicing a strong and well-argued opinion towards acceptance or rejection. PC members tend to be tired of all the hard work, and clear voices are appreciated.

Often, the very last decisions are taken by a vote, e.g. an approval vote where you can vote for as many submissions as you want, and the ones with the most approvals get in. In some rare cases, there might be a physical PC meeting.

\subsection{Physical PC meetings}
Sometimes, the final decisions are taken at a physical PC meeting. Usually, it takes at least two days to go through all borderline cases, and it is very hard (or impossible) to have the same energy and thought level for all these cases. Outcomes of these meetings tend to be (even) more random than votes. At these meetings, you have a \emph{lot} of influence by voicing good arguments at the right moment, it might feel like a political procedure. Of course, it is fun to meet all these clever PC members in person.

\subsection{Polishing reviews}
After the final decisions are taken, the PC chair will ask the PC members to ``polish'' their reviews. In addition, there might be one PC member per submission assigned to summarize the PC discussion for the authors. As a courtesy to the authors, it is useful to reflect some of the PC discussion in the reviews to be sent to the authors. You now had an ``inside view'' of how the decision to accept/reject has taken place, so try to convey the main message to the authors, in particular in case of a ``reject''.

In case of a ``reject'', you should also consider toning down any very enthusiastic reviews in order to prevent the authors from being confused by the conflicting reviews.

\section{After the author notification}\label{sec:wrapup}
After the decisions and reviews have been sent to the authors, most of the hard work is done! There are a few small tasks left before the conference is over.

\begin{mdframed}
During this phase:
\begin{todolist}
\item You help select invited speakers.
\item You help select submissions deserving of an award, longer talk slot, etc.
\item You might chair one or more sessions at the conference.
\end{todolist}
\end{mdframed}

\subsection{Selecting invited speakers and best-(student)-paper awards}
One (pre-)final task of the PC, depending on the conference, is to help select the invited speakers. This process might happen together with input from the steering committee, or local organisers. Some of the accepted submissions might be upgraded to plenary talks.

Often, you can select speakers or results you like, and argue why they should be invited, and sometimes there will be a vote about it. Suggestions for suitable speakers are generally appreciated, as is any thoughtful input (also on issues of diversity). The PC chair should try to balance the topics/areas of invited speakers.

\subsection{Chairing a session at the conference}
PC members are the prime candidates for chairing a session at the conference, as they represent the body of people who have made the selection. You often get to chair the session closest to your own research, which can be fun. A few pointers:
\begin{itemize}
\item Be on time for your session.
\item Make sure you can pronounce all author names (and titles) of your session. If you are unsure, it is OK to double-check with the speaker before the talk starts.
\item Pay attention to the talks.
\item Keep track of the time.
\item Ask the audience for questions, and if there is none (and time), ask at least one question yourself.
\end{itemize}
Although you are welcome to attend the conference (especially if you authored an accepted work), being able to do so is not a requirement for serving on the PC.

\section{Further Resources}
The International Association for Cryptologic Research (IACR) maintains official \href{https://iacr.org/docs/progchair.pdf}{guidelines for PC chairs} and \href{https://iacr.org/docs/reviewer.pdf}{for reviewers} as well as \href{https://iacr.org/docs/}{other documents} relevant to this process.
\begin{description}
\item[2020] Ulle Endriss' \href{https://sites.google.com/view/aamas2021pc/}{FAQ for submissions to AAMAS-2021}
\item[July 2018] Aram Harrow on being QIP PC chair:
\href{https://quantum-aram.github.io/2018/07/17/PC/}{What I learned from being QIP PC chair
}
\item[Aug 2017] Thomas Vidick on chairing the QCrypt PC:
\href{https://mycqstate.wordpress.com/2017/08/07/a-beginners-guide-to-pc-chairing/}{A beginner's guide to PC chairing}
\item[Mar 2015] Boaz Barak on chairing a STOC/FOCS committee: 
\href{https://windowsontheory.org/2015/03/02/tips-for-future-focsstoc-program-chairs/}{Tips for future FOCS/STOC program chairs}
\item[Dec 2008] Graham Cormode on \href{https://citeseerx.ist.psu.edu/viewdoc/download?doi=10.1.1.188.21&rep=rep1&type=pdf}{How NOT to review a paper: the tools and techniques of the adversarial reviewer}
\item[Nov 1989] Ian Parberry on writing a high-quality review: \href{https://jmlr.csail.mit.edu/reviewing-papers/p92-parberry.pdf}{A guide for new referees in theoretical computer science}
\end{description}

\section*{Acknowledgements}
The idea for this guide was conceived during the weekly group meeting of Christian Schaffner's quantum-cryptography group at QuSoft. In autumn 2020, multiple group members were asked to serve on PCs for the first time. While a \href{https://jmlr.csail.mit.edu/reviewing-papers/p92-parberry.pdf}{useful guide} is readily available for first-time referees in TCS, we were unable to find an analogue document for first-time PC members. Reason enough to write our own! Most of the writing was done by CS, most of the editing by YD, all authors have contributed with their experiences and tips. 

We would like to thank Simon Apers, Boaz Barak, Léo Ducas, Ulle Endriss, Aram Harrow, Laura Mančinska, Nikhil Mande, Maris Ozols, and Bart Preneel for their insightful comments and additions. If you have any comments or suggestions, please \href{mailto:c.schaffner@uva.nl?subject=Feedback about PC guide}{let us know} by email to \href{mailto:c.schaffner@uva.nl?subject=Feedback about PC guide}{c.schaffner@uva.nl}. We are happy to incorporate and acknowledge those in future versions of this guide.

\end{document}